\documentclass[conference, a4paper]{IEEEtran} 

\usepackage{cite}
\usepackage[dvips]{graphicx}

\usepackage[cmex10]{amsmath}
\usepackage{amssymb}
\usepackage{algorithmic}
\usepackage{algorithm}
\usepackage{amsfonts}
\usepackage{array}
\usepackage{mdwmath}
\usepackage{mdwtab}
\usepackage{stfloats}
\usepackage{color}

\makeatletter

\newcommand{\Rmnum}[1]{\expandafter\@slowromancap\romannumeral #1@}
\makeatother

\IEEEoverridecommandlockouts

\begin{document}
%
\title{Stochastic Geometry based Interference Analysis of Multiuser mmWave Networks with RIS}
%
%



\author{\IEEEauthorblockN{Joonas Kokkoniemi and Markku Juntti}
\IEEEauthorblockA{Centre for Wireless Communications (CWC)\\
University of Oulu\\
P.O. Box 4500, 90014 Oulu, Finland\\
Email: joonas.kokkoniemi@oulu.fi}}

\maketitle


\begin{abstract}
In this paper, we utilize tools from stochastic geometry to estimate the interference propagation via reconfigurable intelligent surface (RIS) in the millimeter wave (mmWave, 30--300 GHz) band and specifically on the D band (110--170 GHz). The RISs have been of great interest lately to maximize the channel gains in non-line-of-sight (NLOS) communication situations. We derive expressions for stochastic interference level in RIS powered systems and validate those with simulations. It will be shown that the interference levels via RIS link are rather small compared to the designed RIS link or the LOS interference as the random interference loses significant part of the RIS gain. We also analyse the validity of far field channel and antenna gains in the near field of a large array. It is shown that, while the high frequency systems require large arrays that push the far field far away from the antenna, the far field equations are very accurate up to about half way of the near field.
\end{abstract}


%
\IEEEpeerreviewmaketitle


\section{Introduction}

The fifth generation (5G) and beyond networks seek higher data rates in the millimeter wave (mmWave, 30--300 GHz) and terahertz frequency bands (300 GHz--10 THz) \cite{Akyildiz2014,Latva-Aho2019}. These high frequency bands offer far higher bandwidths than the traditional ultra high frequency bands (UHF, 300 MHz--3 GHz). This allows very high theoretical data rates even with low spectral efficiency transmission methods, such as by using error tolerant low modulation orders. However, while very appealing for the various high data rate applications, such as backhauling, data kiosks, and augmented reality, the high frequency communications suffers from very high channel losses.

Future high frequency high bandwidth systems rely on low loss channels and high gain antennas for efficient operation. Reconfigurable intelligent surfaces (RISs) are seen as potential technologies to intelligently modify the propagation environment in order to maximize the channel gain \cite{DiRenzo2020}. The RISs theoretically offer great benefits in the situations where LOS path is obstructed by in practice reflecting the incoming waves into wanted direction. In the other words, the RIS phases are manipulated to offer high reflection power towards wanted directions, often in anomalous directions as compared to normal reflections. There is a good collection of details on different RIS structures and latest research on RISs by Di Renzo \textit{et al.} \cite{DiRenzo2020}.

Most of the RIS control algorithms in the literature focus on optimizing the performance of a single link aided by a RIS. In such a case, the signals from other transmitters appear as interference at the receiver. The impact of such interference has not been widely studied yet, but some works on multiuser RIS control has been conducted \cite{9153884-Zhang2020}. Therein a scheduled access and multiuser RIS control based system optimization has been considered.

Stochastic geometry has been shown to provide an efficient set of tools for network performance evaluations \cite{Haenggi2009a,Haenggi2009,ElSawy2013,Kokkoniemi2018a,Kokkoniemi2017,Kokkoniemi2016d,Kokkoniemi2020a}. Traditionally, network performance has been studied with simulation models. Those remain relevant today as they allow replication of system parameters in computer simulation platform for testing and overall system evaluation. Stochastic geometry relies on mathematical representation of the system and stochastic average values for, e.g., the antenna gains, user densities, etc. Thus, the total response of the system can be achieved by geometrically integrating over the system. This provides a fast way to overcome complex and time consuming simulation models. However, simulation models offer useful tools to study system performance on signal level to test, e.g., signal models or beamformers, or to deploy and study complicated environments, e.g., via ray tracing simulations. The stochastic tools are more useful in studying general power distributions in relatively simple environments.

In this paper, we utilize stochastic geometry for estimation of the interference propagation via RIS. The frequency band of interest herein is the D band (110--170 GHz) as this is among the frequency band for next generation systems as well as the main interest of European H2020 project ARIADNE \cite{ARIADNED11}. The basic assumptions herein are that there is a desired RIS link between Tx and Rx, and interfering Txs with no attempt to perform multiuser RIS control. The particular interest is to model the interference behavior at RIS and subsequently the response of the interference at the desired Rx. The focus of this work is to model the interference on general level. In the future work, we will extend the work herein to more sophisticated use cases with multiuser control at RIS. We further develop a simulation model to validate the accuracy of the derived stochastic models. The simulation validation is partially due to the path loss and antenna models are meant for plane wave propagation, that is, for far field propagation. The high path losses in the mmWave and beyond bands force to utilize large antenna arrays to provide enough antenna gain to overcome the channel losses. This means that the near field of the antenna arrays are also pushed far away from the antenna with respect to the wavelength. The simulation results show that the derived stochastic models are 1) accurate, 2) work relatively well in the near field of the RIS as well. It is shown that the RIS interference is not as large as the direct LOS interference. There are some papers on stochastic geometry modelling of RIS systems, e.g., \cite{Kishk2021,DiRenzo2019,Hou2021,Lyu2020}. However, to the best of our knowledge, there are no papers on interference propagation via RISs.

The rest of this paper has been organized as follows. The system model is given in Sec. \ref{sec:2}, the derived stochastic interference models are given in Sec. \ref{sec:3}, the simulation model used for validation is briefly presented in Sec. \ref{sec:4}, numerical results and validations are given in Sec. \ref{sec:5}, and finally the conclusions are drawn in Sec. \ref{sec:6}.


\section{System Model}
\label{sec:2}

\begin{figure}[t!]
    \centering
    \includegraphics[width=3.2in]{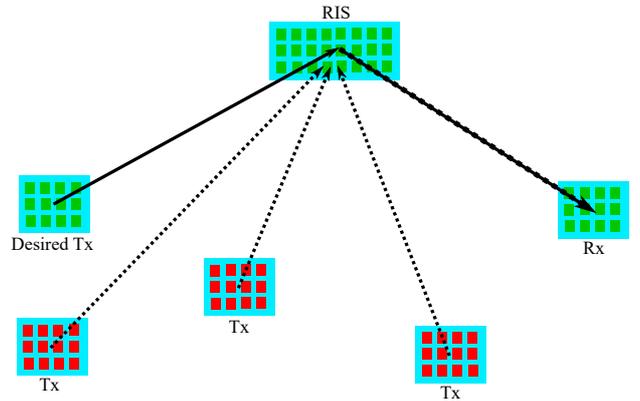}
    \caption{System model illustration.}
    \label{fig:sys}
\end{figure}

The system model utilized herein is given in Fig.\ \ref{fig:sys} and it is formed of the desired Tx and Rx, RIS, and random interfering Txs. The RIS phases, as well as the Tx and Rx beamformers (LOS, Tx--RIS, and RIS--Rx beamformers) are calculated based on the positions of the three main system entities. The RIS always beamforms between the desired Tx and Rx. On the other hand, the random interfering Txs are assumed to be randomly distributed about the RIS. More details on the interfering Txs is given in the next section, but those beamform either randomly or towards the RIS.

All the system entities utilize planar antenna arrays. We assume isotropic elements at $\lambda/2$ spacing in both horizontal and vertical directions, where $\lambda$ is the wavelength of the system. The center frequency is assumed to be 140 GHz. This leads to very compact antenna arrays. Especially in the case of RIS, the small physical area at high frequencies means small intercepted energy, and thereafter leads to need of large arrays to 1) capture enough energy, and 2) to produce enough gain towards desired directions. The need for large antenna arrays in the high frequencies is universally true for all the wireless network elements due to large path losses at mmWave and above frequency bands.

Since the stochastic models for the signals given below are derived by assuming far field communications, the main point herein is to analyse the system performance in the far field regime. However, as the RISs can potentially be very large, there is also a great chance that the served nodes are in the near field of it. The near field region of an antenna array reaches to about
\begin{equation}
    R_\text{NF} = \frac{2D^2}{\lambda},
\end{equation}
where $R_\text{NF}$ is the radius of the near field region around the array and $D$ is the maximum dimension of the antenna array. It will be shown by simulations that the far field assumptions for the stochastic models to be accurate in the near field up to about $D^2/\lambda$ distance away from the RIS (or any other array).

\section{Interference Characterization with RISs}
\label{sec:3}

We assume LOS channels on all links. Furthermore, as mentioned above, we assume far field in all derivations. This assumptions gives an easy way to model the system and it is also very accurate deep into the near field as it will be shown in the numerical results. The general high frequency LOS link gain, including the transmit power, path loss, and antenna gains, is obtained as \cite{Jornet2011}
\begin{equation}
\label{eq:LOSresp}
h_\text{LOS}(r,f) = P_\text{Tx}G_\text{Tx}G_\text{Rx}\frac{c^2\exp(-\kappa_\text{a}(f)r)}{(4\pi r f)^2},
\end{equation}
where $P_\text{Tx}$ is the transmit power, $\kappa_\text{a}(f)$ is the absorption coefficient at frequency $f$ (given we are close to absorption line introducing additional loss), $r$ is the distance from Tx to Rx, $c$ is the speed of light, and $G_\text{Tx}$ and $G_\text{Rx}$ are the antenna gains. Having the base signal model defined, we can next look into the interference power.

The aggregate interference at Rx can be analytically evaluated as~\cite{Kokkoniemi2018a,Kokkoniemi2017,Kokkoniemi2016d,Kokkoniemi2020a}
\begin{equation}
\label{eq:Iaggr}
I_\text{aggr} = \sum\limits_{i\in \Phi} h(r_i,f),
\end{equation}
where $h(r_i,f)$ is the path gain model including the expected antenna gains and transmit powers. The summation is done over an active set of interferers $Phi$. The path gain model depends on the link type. The LOS interference gain is obtained as
\begin{equation}
h_\text{I,LOS}(r_i,f) = P_\text{Tx}\frac{c^2\exp(-\kappa_\text{a}(f)r)}{(4\pi r f)^2},
\end{equation}
assuming all the nodes share the same transmit powers and other signal properties. This expression is true for nodes that transmit in random directions. If there is directionality in the links, the appropriate expected gains should be added. For instance, if all the nodes are randomly distributed about the Rx, but are pointed at the Rx, the expected Tx gains would be G$_\text{Tx}$ and the expected Rx gain would remain unity.

With RIS, the situation changes a bit as we assume that the RIS array is an ideal phase shifter and reflector. Therefore, the electric size of an antenna $c^2/4\pi$ in above is replaced by the area of the RIS $A_\text{RIS}$. This is the capture area for the energy at the RIS and we assume it to be $(N_\text{RIS,x}-1)d\times(N_\text{RIS,y}-1)d$, where $N_\text{RIS,x}$ and $N_\text{RIS,y}$ are the number of horizontal and vertical antenna elements, respectively, and $d (=\lambda/2)$ is the antenna element separation. In the numerical results, we assume square RIS. This structure has the advantage of the smallest near field distance due to its maximum dimension is smallest among all rectangular shapes.

The RIS desired link gain is obtained as
\begin{equation}
\label{eq:RISresp}
h_\text{RIS} = P_\text{Tx} G_\text{Tx}G_\text{Rx}G_\text{RIS}\frac{\exp(-\kappa_a (r_\text{Tx,RIS}+r_\text{RIS,Rx}))c^2A_\text{RIS}}{(4\pi)^3f^2r_\text{Tx,RIS}^2r_\text{RIS,Rx}^2} ,
\end{equation}
where $r_{Tx,RIS}$ and $r_{RIS,Rx}$ are the Tx--RIS and RIS--Rx distances, $A_\text{RIS}$ is the RIS area (or capture area for the incoming energy) and $G_\text{RIS}$ is the total gain of the RIS. The the channel gains from the interfering nodes to RIS become
\begin{equation}
h_\text{I,RIS}(r,f) = \frac{A_\text{RIS}\exp(-\kappa_\text{a}(f)r)}{4\pi r^2},
\end{equation}
or
\begin{equation}
h_\text{I,RIS}(r,f) = A_\text{RIS}G_\text{Tx}\frac{\exp(-\kappa_\text{a}(f)r)}{4\pi r^2},
\end{equation}
if the interfering users are pointed towards the RIS.

Given these channel gains and the expression for the aggregate interference in \eqref{eq:Iaggr}, we can derive the aggregated interference levels similarly as we did in past works ~\cite{Kokkoniemi2018a,Kokkoniemi2017,Kokkoniemi2016d,Kokkoniemi2020a}. Otherwise the assumptions in modelling are as follows. We assume ALOHA transmissions. This simplifies the analysis, albeit giving the worst case interference due to no coordination in the transmissions. We assume that the users are Poisson distributed and all the interfering users have the same transmit power and antenna configurations. We assume that all the antennas are planar antennas, including the RIS, which is a planar reflective array. Single antenna elements are assumed to be isotropic. Hence, the total maximum antenna in the far field is equivalent to the number of antenna elements ($G_\text{Tx/Rx} = N_\text{Tx/Rx}$, where $N_\text{Tx/Rx}$ is the total number of antenna elements at Tx/Rx).

The desired LOS response is directly given by \eqref{eq:LOSresp}) and the received power via desired RIS link is given by \eqref{eq:RISresp}. If the users are dropped on 2D plane, the aggregated LOS interference (i.e., direct interference without RIS) at Rx becomes
\begin{equation}
\mathbb{E}[I_\text{Rx}] = \frac{P_\text{Tx} c^2 p_T p_N\lambda_U}{8\pi f^2}  \int\limits_r \frac{\exp(-\kappa_a r)}{r}dr,
\end{equation}
where $\lambda_U$ is the user density, $p_T$ is the transmit probability of a Tx, and $p_N$ is the non-blocked probability of the path between Tx and Rx. These latter two terms act as thinning operators for the Poisson distributed number of users or the density of users. These terms are utilized in the future work on the analysis of the blocking probability on the RIS link performance and are assumed to be unity in this work.

The interference level at the RIS becomes
\begin{equation}
\mathbb{E}[I_\text{RIS}] = \frac{P_\text{Tx}A_\text{RIS} p_T p_N\lambda_U}{4}  \int\limits_r \frac{\exp(-\kappa_a r)}{r}dr,
\end{equation}
or
\begin{equation}
\mathbb{E}[I_\text{RIS}] = \frac{P_\text{Tx}G_\text{Tx}A_\text{RIS} p_T p_N\lambda_U}{4}  \int\limits_r \frac{\exp(-\kappa_a r)}{r}dr,
\end{equation}
if the interfering users beamform towards the RIS. These equations take into account that the RIS is mounted on a wall and the users are distributed on a semicircle around the RIS. The corresponding RIS--Rx link is the same in all cases, and the total interference level at Rx is
\begin{equation}
\mathbb{E}[I_\text{RIS,Rx}] = \mathbb{E}[I_\text{RIS}]\frac{c^2 G_\text{Rx}\sqrt{G_\text{RIS}}}{(4 \pi r_{RIS,Rx} f)^2}.
\end{equation}
More detailed derivations are given in the future work, but the numerical results show that these models are accurate with respect to the simulation model. The simulation model itself is briefly described in the next section.


\section{Simulation Model}
\label{sec:4}

The simulation model was developed to test the validity of the RIS link models. Whereas the theoretical models rely on stochastic gain values and integrations over the space to obtain the average interference levels, the simulation model was built from ground up to model the exact phase behavior of the system. Every network element including each antenna element has physicalized location. Based on the locations, inter-element distances, and angles between the network elements (Rx, RIS, and Txs), the exact phases of the antenna elements are calculated with linear planar antenna array factors \cite{Balanis2016}. After all the phases have been calculated, the response every single antenna element to all the target antenna element are calculated. By calculating all the responses based on the phases, the simulator gives the exact response regardless of near-field or far field, as long as the Txs and Rx are in the far field of a single antenna element.

In the simulation model, for instance, for the interfering nodes, the amplitude response of a single interfering Tx at RIS element ($k,l$) is
\begin{equation}
\begin{aligned}
    A_\text{RIS}(k,l) &= \sqrt{\frac{P_\text{Tx}A_\text{RIS}}{4\pi}}\sum\limits_{i=1}^{N_\text{Tx,el}}\sum\limits_{j=1}^{N_\text{Tx,az}} \\&\Xi_\text{RIS}(k,l)\xi_\text{Tx}(i,j) \frac{e^{-\frac{1}{2}\kappa_a r_{i,j\rightarrow k,l}}e^{-j2\pi f r_{i,j\rightarrow k,l}/c}}{r_{i,j\rightarrow k,l}},
\end{aligned}
\end{equation}
where $N_\text{Tx,el}$, $N_\text{Tx,az}$, $N_\text{Rx,el}$, and $N_\text{Rx,az}$ are the number of Tx and Rx antenna elements in elevation and azimuth directions (vertical/horizontal antenna), $r_{i,j\rightarrow k,l}$ is the distance from Tx antenna element ($i,j$) to RIS element ($k,l$), $\exp(-j2\pi f r_{i,j\rightarrow k,l}/c)$ is the linear phase shift, and $\Xi_\text{RIS}(k,l)$ and $\xi_\text{Tx}(i,j)$ are the phases of the RIS and Tx elements, respectively. Then the total received power from a single Tx at the Rx becomes
\begin{equation}
\begin{aligned}
    P_\text{Rx} &= \Bigg|\sum\limits_{i=1}^{N_\text{Rx,el}}\sum\limits_{j=1}^{N_\text{Rx,az}}\sum\limits_{k=1}^{N_\text{RIS,el}}\sum\limits_{l=1}^{N_\text{RIS,el}} \\& A_\text{RIS}(k,l)\zeta_\text{Rx}(i,j) \frac{c e^{-\frac{1}{2}\kappa_a r_{k,l\rightarrow i,j}}e^{-j2\pi f r_{k,l\rightarrow i,j}/c}}{4\pi r_{k,l\rightarrow i,j} f }\Bigg|^2,
\end{aligned}
\end{equation}
where $r_{k,l\rightarrow i,j}$ is the distance from the RIS element ($k,l$) to Rx element ($i,j$) and $\zeta_\text{Rx}(i,j)$ are the Rx antenna phases. Similarly, by summing the phases, the exact response of any Tx to Rx, direct or via RIS, can be calculated. This simulation model gives valuable information on the limits of the theoretical models, but also allows in the future work to test the impact of different beamforming algorithms on the system performance, or to study the impact of the phase noise on the RIS beamforming, to mention a couple of options.

To achieve comparable setting to the derived stochastic models, the user locations are Poisson distributed. The RIS is assumed to be mounted on a wall, and the users are distributed in semicircle around the RIS. All the users in these simulations transmit constantly. However, the number of users is thinned based on the transmit and blocking probabilities similarly as in the stochastic models if those are modelled. The RIS is assumed to be a perfect reflector, or on the other hand, an ideal phase sifter. Hence, the energy intercept area is calculated similarly as in the stochastic models from the number of antenna elements. Lastly, all the Tx are randomly oriented. If the Tx is pointed at the RIS, the Tx beamformer is calculated towards the RIS, otherwise the Tx beamforms towards the randomly picked direction (azimuth and elevation).


\section{Numerical Results}
\label{sec:5}

In this section, we study performance of the derived models, as well as look into the interference levels caused by the RIS on the desired RIS--Rx link. Across all the results, the center frequency is 140 GHz and transmit power is 1 W for all Txs. One important aspect is to evaluate the accuracy of the models in the near field of the arrays. When the frequencies grow, the required antenna gains also increase. This means large near field radius around the antennas. The traditional propagation and antenna gains (e.g., the array factor) are based on plane wave propagation, i.e., far field propagation. Thus, we need to understand the limits where the far field equations are valid with very large antenna arrays. Those are looked into first before validation of the interference models and the results on the interference levels in RIS powered systems.

\subsection{Accuracy of the Far Field Assumption}

\begin{figure}[t!]
    \centering
    \includegraphics[width=3.3in]{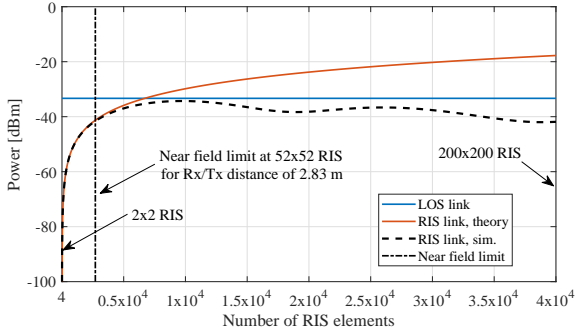}
    \caption{Performance of the direct RIS link versus simulation for large RIS sizes where Rx and Tx are close or in the near field of the RIS.}
    \label{fig:LOS}
\end{figure}

\begin{figure}[t!]
    \centering
    \includegraphics[width=3.3in]{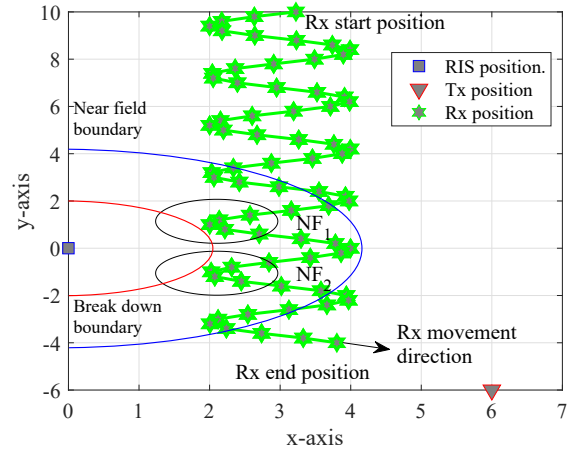}
    \caption{The locations of Tx and RIS, and the Rx movement pattern in the simulation for moving Rx. The axes units are given in meters.}
    \label{fig:MRx1}
\end{figure}

\begin{figure}[t!]
    \centering
    \includegraphics[width=3.3in]{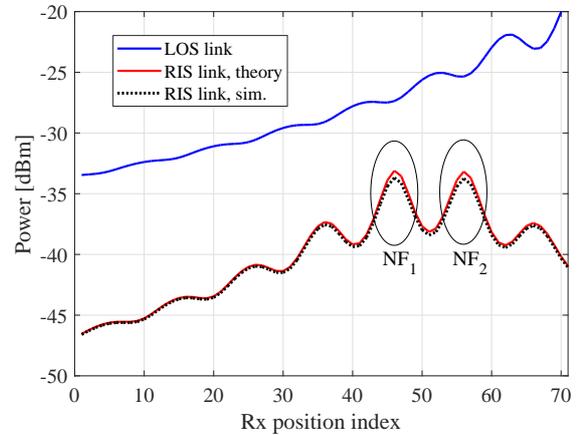}
    \caption{Received power at moving Rx including the near field points where the far field assumption starts to give inaccurate response.}
    \label{fig:MRx2}
\end{figure}

In Fig.\ \ref{fig:LOS}, receiver power of the desired link via RIS is given for by the theoretical model and by the simulations. For comparison, the direct LOS power is also given. This figure gives the sanity check for the validity of the far field assumptions for the theoretical models. The RIS size is varied from 2-by-2 to 200-by-200. Thus, the numbers of RIS elements in the higher end are quite unrealistic. However, the figure shows that the far field assumptions work quite well up to about $D^2/\lambda$. For instance, this figure was calculated for the Tx and Rx at 2.83 m away from the RIS. With $\lambda/2$ element separation and square RIS, the Rx the Tx are in the near field if RIS is larger than 52-by-52. When going closer in the near field than about $D^2/\lambda$, the theoretical far field based models no longer give reliable results. In fact, we can see that simulation model does not exceed the LOS response at any RIS size. This is due to the spherical signal propagation in the near field of the array, and the fact that the RIS link is slightly longer that the direct LOS path (5.66 meters versus 4 meters for the direct LOS path). It would be possible to achieve higher near field response by aligning the phases at the RIS, i.e., by doing focusing in the near field. However, this also requires one more degree of freedom as the beamforming algorithm should know the exact positions of the Tx(s) and Rx(s).

To further check the validity of the far field assumptions in the near field, we calculated the Rx power of a moving Rx. The movement trajectory and the RIS and Tx locations are shown in Fig.\ \ref{fig:MRx1}. The Rx power for direct LOS link as well for the RIS link with simulation and the theory is given in Fig.\ \ref{fig:MRx2} and those were calculated assuming 16-by-16 antenna array at Tx, 64-by-64 RIS, and the Rx had 32-by-32 element antenna array. Large arrays were utilized so that the near field of the RIS would be large. With this RIS size, the near field boundary reached to about 4.2 meters away from the RIS (shown in Fig.\ \ref{fig:MRx1}). The break down boundary in Fig.\ \ref{fig:MRx1} is $D^2/\lambda$, or about 2.1 meters away from the RIS. This is exactly half of the formal near field boundary. We can see that the far field equations estimate the correct Rx power very accurately up to about this break down boundary. The two ellipsoids, NF$_1$ and NF$_2$ in Figs. \ref{fig:MRx1} and \ref{fig:MRx2} show that close to this boundary, the far field assumption begins to fail. Thus, this break down boundary is as close to array as where the far field equations produce feasibly accurate results (depending on the accuracy requirement of the application). Furthermore, we can see that the LOS response is always clearly better than the RIS link. This is, however, expected result as the RISs are in general seen the most beneficial when the LOS link is not available. The RIS link is also always longer than the LOS link, which in the presence of high path loss gives the advantage for the pure LOS path.

\subsection{Validation of the Stochastic Interference Models}

\begin{figure}[t!]
    \centering
    \includegraphics[width=3.3in]{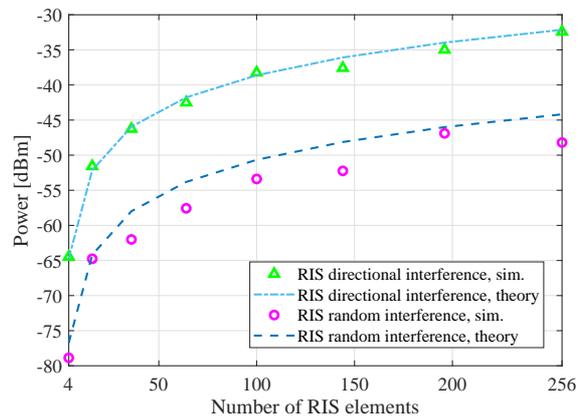}
    \caption{Benchmark of the theory versus simulation for the interference at Rx via RIS from random Txs pointed at RIS and from random Txs pointed at random directions.}
    \label{fig:sim}
\end{figure}

Fig.\ \ref{fig:sim} shows the performance of the stochastic RIS interference models to the simulation results. The RIS size was varied and the Txs and the Rx had 4-by-4 antenna arrays. We see very good match in the simulation model versus the theory in the case when the random Txs beamform towards the RIS. This is attributed to less randomness in the network and shows that the stochastic model is correct. The challenge with the simulation model is to achieve enough data in reasonable time whereas the stochastic model gives the output in closed form. The number of possible phase combinations between all the antenna elements is very large, especially if the nodes transmit to random directions. For instance, these simulations were run for relatively modest size RISs with 4000 users randomly distributed only at most 2 meters away from the RIS. Then the simulations were repeated over 1000 realizations. Yet, the random direction simulations do not converge as it is very unlikely that the phases would align perfectly to give the maximum gain. This is far more likely if the Txs are pointed towards the RIS as this figure shows. In any case, the result herein shows that the stochastic models are correct.

\subsection{Interference in RIS Systems}

\begin{figure}[t!]
    \centering
    \includegraphics[width=3.3in]{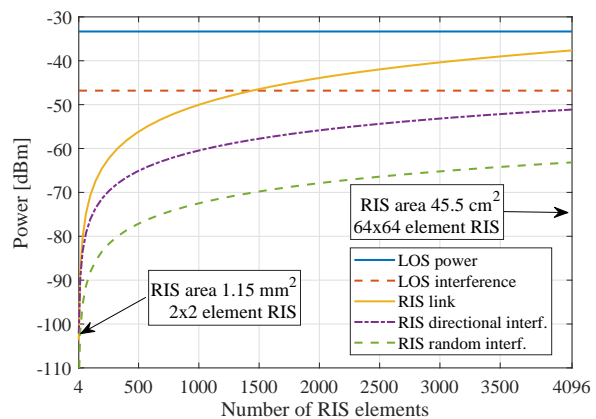}
    \caption{LOS power versus stochastic RIS links, as well as LOS interference power for comparison.}
    \label{fig:res}
\end{figure}

Lastly, in Fig.\ \ref{fig:res} we give the purely theoretical response of the RIS links (desired plus interference) at the Rx. For comparison, the direct LOS path and LOS interference are also given. In this case, the RIS size is varied from 2-by-2 to 64-by-64. The average number of interfering users was set to 10 with maximum distance of 10 meters from the RIS. The Tx and Rx are at 2.83 m away from the RIS. There are few interesting takeaways from this figure. Firstly, as the 140 GHz makes the antenna elements physically small with relatively large path loss, the capture area of RIS is also very small. These translate into small amount of captured energy for the redirection towards the Rx. Therefore, the RIS needs to be sufficiently large to provide enough gain for good received power levels. The more significant finding is that the interference levels via RIS are rather modest compared to the desired RIS link. The main reason is that the interference in average loses the most of the RIS gain as the RIS is aligned between the desired Tx and Rx. The directional interference does offer relatively high interference level, but we have to remember that the ALOHA scheme herein gives the worst case interference. The random interference, on the other hand, loses quite a bit more as it in average also loses the Tx antenna gains. This is obviously a good thing for the RIS systems, as the highest RIS gain is only achieved when the RIS phases are aligned with the Rx and Tx phases. Thus, the designed RIS link always has a significant advantage over the interfering links. Obviously, with high enough numbers of simultaneously transmitting Txs, the interference level can catch up with the desired link, but more significant problem is the direct LOS interference as shown in Fig.\ \ref{fig:res}.


\section{Conclusion}
\label{sec:6}

In this paper, we studied the interference propagation via RIS. Stochastic models for the RIS interference were derived and those were shown to be accurate by simulations. In the numerical results, it was shown that the RIS interference is not a significant problem due to interference via RIS suffers from reduced gain in comparison to the designed RIS link. Far higher interference levels are caused by the direct LOS interference. However, these are subject to the exact system settings, such as the LOS probability. The impact of the LOS probability on the average interference level as well as on the desired link performance are looked in the future work.

\section*{Acknowledgement}

This work was supported by the Horizon 2020, European Union's Framework Programme for Research and Innovation, under grant agreement no.~871464 (ARIADNE). It was also supported in part by the Academy of Finland 6Genesis Flagship under grant no.~318927.

%


\ifCLASSOPTIONcaptionsoff
  \newpage
\fi

\bibliographystyle{IEEEtran}
\bibliography{main.bbl}

\end{document}